\documentclass[11pt,twoside]{article}
\usepackage{asp2004}
\usepackage{psfig}

\begin{document} 
\title{Luminosity Profiles of Resolved Young Massive Clusters} 

\author{Fran\c cois Schweizer}
\affil{Carnegie Observatories, 813 Santa Barbara Street, Pasadena, \\
CA 91101-1292, USA}

\begin{abstract} 
Young massive clusters differ markedly from old globular clusters in
featuring extended, rather than tidally truncated, envelopes.  Their
projected-luminosity profiles are well fit by Elson-Fall-Freeman models
with core radii $0.3\ {\rm pc} \la r_{\rm c}\la 8$~pc and {\it power-law
envelopes} of negative exponent $2.0\la \gamma\la 3.8$.  These envelopes
form within
the first few 10$^6$\,yr and last $\sim$10$^8$\,--\,10$^{9.5}$\,yr,
depending on the environment. Many YMCs show clumpy substructure that
may accelerate their initial relaxation.  The cores of Magellanic-Cloud
clusters show universal expansion from $r_{\rm c}< 1$~pc at birth
to $r_{\rm c}= 2$\,--\,3~pc after 10$^8$\,yr, but then seem to evolve
along two bifurcating branches in a $r_{\rm c}$\,--\,log(age) diagram.
The lower branch can be explained by mass-loss driven core expansion
during the first 10$^9$\,yr, followed by slow core contraction and the
onset of core collapse due to evaporation.  The upper branch, which
shows continued core expansion proportional to logarithmic age, remains
unexplained.  There is strong evidence for rapid mass segregation in
young clusters,
yet little evidence for top-heavy IMFs or primordial mass segregation.
Finally, YMCs show similar structure throughout the Local Group and as
far away as we can resolve them ($\la$20 Mpc).
\end{abstract}

\section{Introduction}

Luminosity profiles have long been used to study the structure and
evolution of old globular clusters (e.g., King 1966;  Illingworth \&
Illingworth 1976).
Their application to the study of young massive clusters is of more
recent origin, having begun in earnest with the classic study of
ten rich, 8\,--\,300~Myr old clusters in the LMC by Elson, Fall, \&
Freeman (1987, hereafter EFF87).
As this study demonstrated, the luminosity profiles of YMCs differ
significantly and systematically from those of old globulars.  The
observed differences contain important clues concerning the formation
and dynamical evolution of massive star clusters.

Among the various motivations for studying luminosity profiles (hereafter
LPs) of YMCs is our desire to understand the effects of mass segregation
as a function of time and to separate any possible primordial
mass segregation from the longer-term segregation caused by energy
equipartition.  Young massive clusters offer two main advantages:
They still contain stars over most of the mass range of $\sim$0.1\,--\,120
$M_{\sun}$, and they allow us to trace evolutionary effects directly
as a function of cluster age.

Luminosity profiles can also yield information on core collapse,
both slow and fast (``equipartition instability,'' Spitzer 1969),
and on the possible formation of central black holes in clusters.
Finally, such profiles can help us evaluate the impact of mass loss
on cluster formation and disruption, yielding estimates of the fraction
of field stars that may have originated in clusters.

\section{Luminosity Profiles: Basics}

We are all familiar with King (1966) model profiles, shown in
Fig.~\ref{fig1}a, which fit the LPs of old globular clusters better than
Gaussian or modified Hubble profiles do.  King profiles are derived
from model clusters with equal-mass stars and truncated Maxwellian
velocity distributions (``lowered isothermal models''), and are
characterized by three parameters: $I_0$, $r_{\rm c}$, and $r_{\rm t}$
(central surface brightness, core radius, and tidal radius).  The
concentration index, $c \equiv \log r_{\rm c}/r_{\rm t}$, is a measure
of how strongly tidal truncation has affected the cluster.

\setcounter{figure}{0}
\begin{figure}[t]
\centerline{\hbox{\vsize 6.0truecm
\psfig{figure=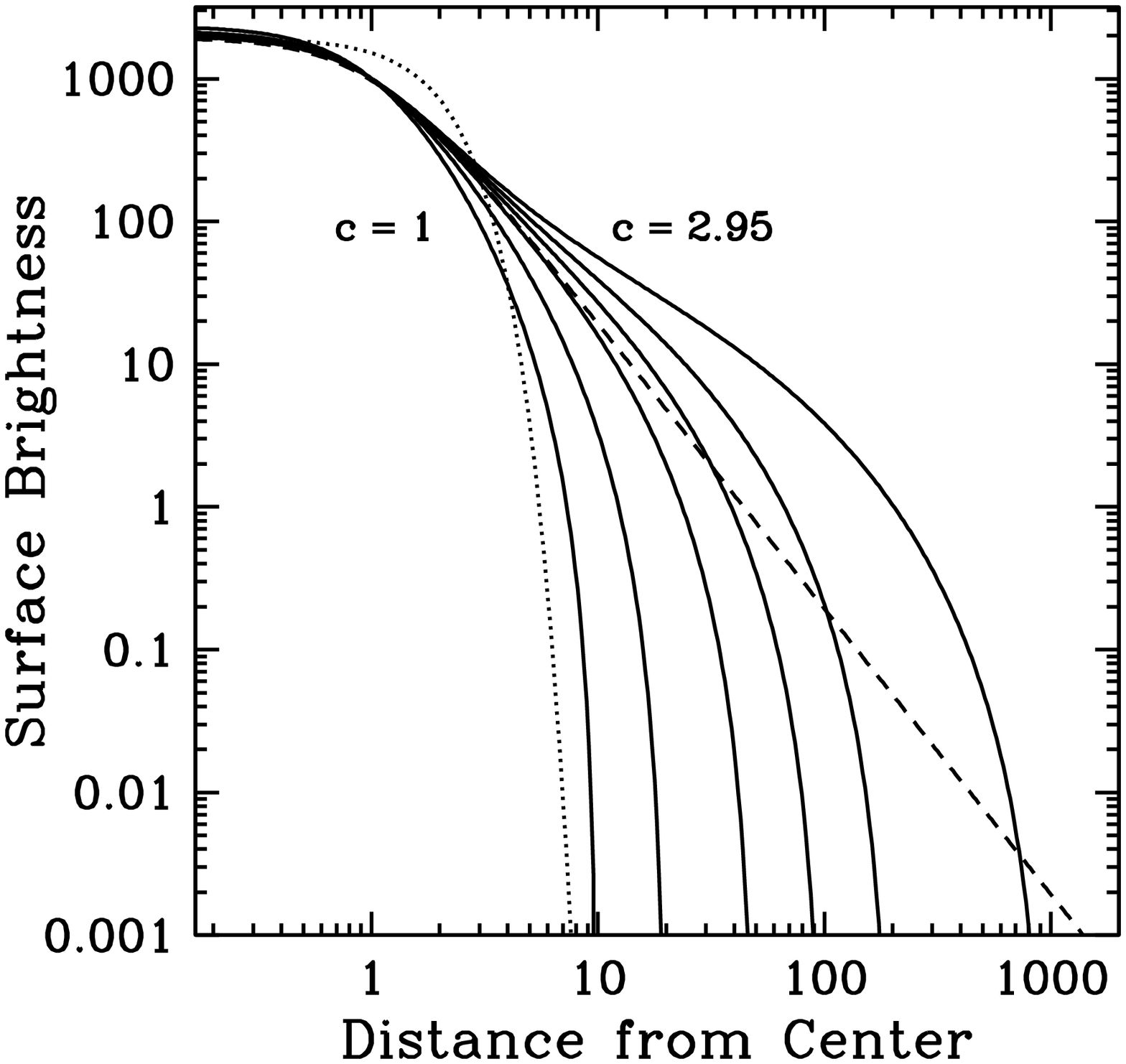,height=5.5cm}
\hskip 0.2truecm
\psfig{figure=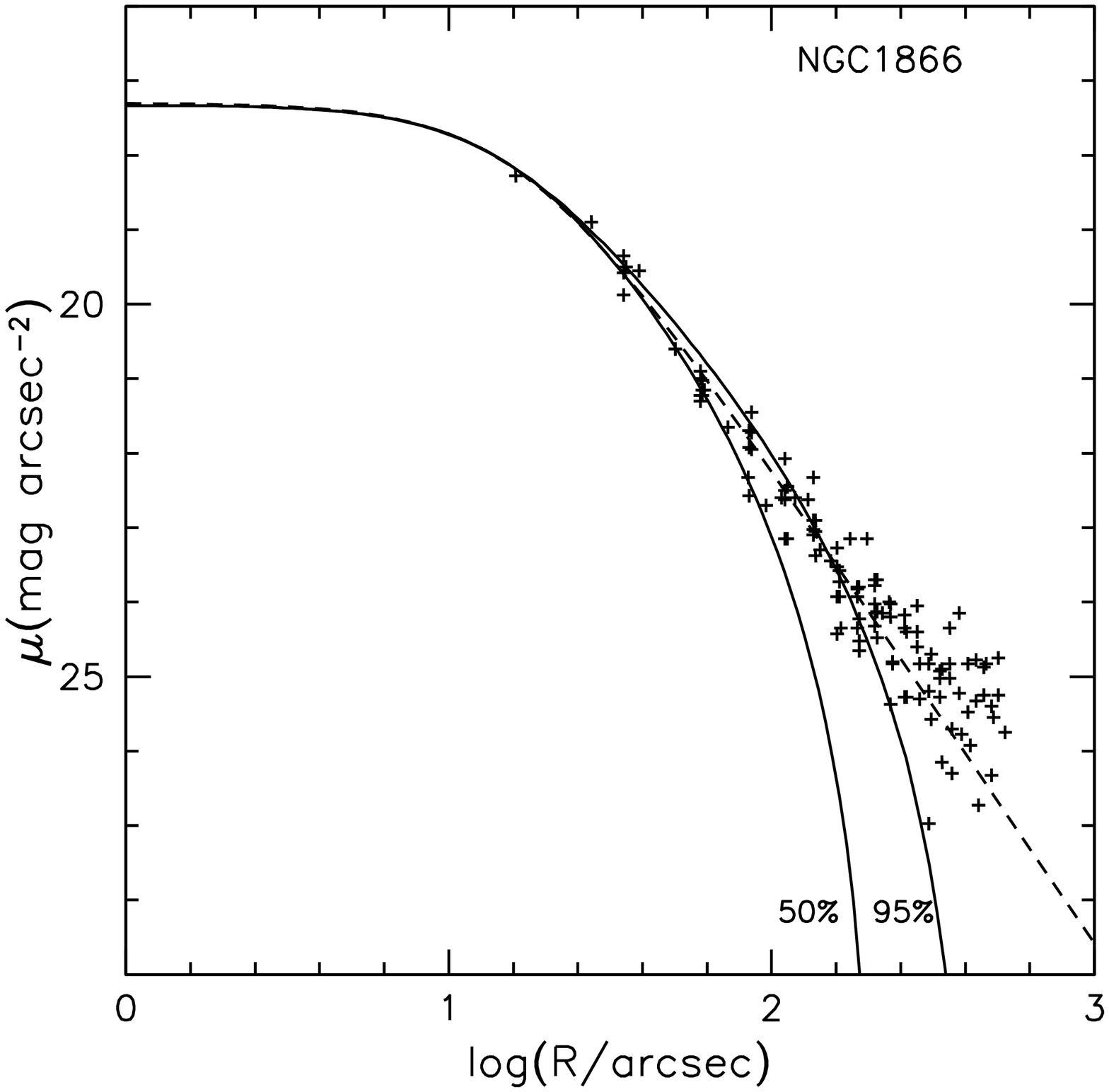,height=5.5cm}
}}
\caption{
(a) ({\it left}) Family of King profiles ({\it solid lines}, $c =
1$\,--\,2.95) compared with Gaussian ({\it dotted line}) and modified
Hubble profile ({\it dashed}, from Carlson \& Holtzman 2001). King
profiles fit old globular clusters well.
(b) ({\it right}) Surface-brightness measurements for 130~Myr-old
LMC cluster NGC~1866 ({\it data points}) compared with King profiles
({\it solid curves}) and best-fit EFF profile ({\it dashed}).  The EFF
profile fits the young-cluster profile with its power-law envelope best
(from Lupton et al.\ 1989).
}
\label{fig1}
\end{figure}

As EFF87 first showed and Elson (1991) further demonstrated, young
massive clusters in the LMC feature
power-law envelopes and are well fit by model profiles of the form
$I(r) = I_0 (1 + r^2/a^2)^{-\gamma/2}$, much better so than by King
profiles.  At radii $r \gg a$, these EFF profiles take the simple
power-law form $I(r) \sim r^{-\gamma}$. Like the King profiles, they
are characterized by three parameters: $I_0$, $a$, and $\gamma$ (central
surface brightness, characteristic radius, and power-law slope).  The
characteristic radius $a$ is related to the core
(= half-central-surface-brightness) radius $r_{\rm c}$ through
$r_{\rm c} = a (2^{2/\gamma}-1)^{1/2}$.  Figure~\ref{fig1}b illustrates
that the 130~Myr-old LMC cluster NGC 1866 is, indeed, better fit by
an EFF model profile than by King profiles (EFF87; Lupton et al.\ 1989).

In trying to understand cluster dynamics, three characteristic time
scales are important (e.g., Spitzer 1987; Meylan 2003): (1) the
crossing time $t_{\rm cr}$ during which a star typically crosses the
system; (2) the half-mass relaxation time $t_{\rm rh}$ during which
stellar encounters redistribute energies to the point of setting up a
near-Maxwellian velocity distribution within the half-mass radius; and
(3) the evolution time $t_{\rm ev}$, which is the time it takes for
slow dynamical processes like the evaporation of stars from the
cluster to significantly change the cluster size and profile. It
has long been known that for globular clusters\ \
$t_{\rm cr} \ll t_{\rm rh} \ll t_{\rm ev}$, with typical values of
$t_{\rm cr} \approx 10^6$ yr, $t_{\rm rh} \approx 10^8$ yr, and
$t_{\rm ev} \approx 10^{10}$ yr.  Thus, the LPs of young massive
clusters should yield valuable information about equipartition and
relaxation processes in the inner parts of these clusters.

\begin{figure}[t]
\centerline{\hbox{
\psfig{figure=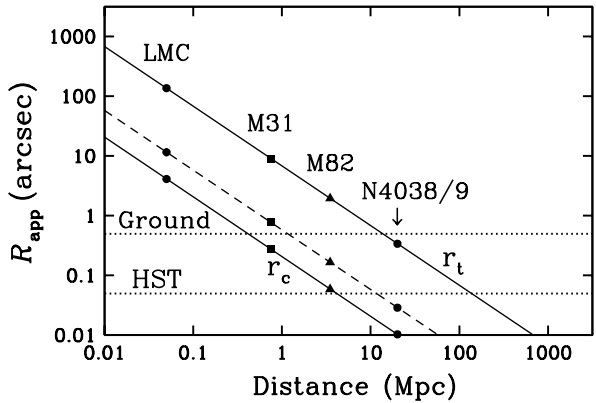,height=5.3cm}
}}
\caption{
Apparent radii of a typical Milky-Way globular cluster at different
hypothetical distances.  Sloping lines mark the radii $r_{\rm c}$,
$r_{\rm eff}$ ({\it dashed}), and $r_{\rm t}$.
}
\label{fig2}
\end{figure}

How far can we hope to study LPs of clusters in sufficient detail
to learn about the structural details that interest us?
Figure~\ref{fig2} shows the {\it apparent} sizes of the core, half-light,
and tidal radius of a typical Milky-Way globular cluster as a function
of its hypothetical distance in Mpc.  This typical globular, with radii
taken as the
median values of 143 globulars listed by Djorgovski (1993), has
$r_{\rm c} = 1.0$ pc, $r_{\rm eff} = 2.8$ pc, and $r_{\rm t} = 33$ pc,
and a concentration index of $c = 1.53$.  As the three sloping lines
of Fig.~\ref{fig2} indicate, this cluster---if placed at the
distance of the LMC---would show a fully resolved core even from the
ground (assumed seeing of $0\farcs5$ FWHM).  At the distance of M31
the same cluster would still appear well resolved with {\it HST}, while at
the 3.5 Mpc distance of M82 its core would be only marginally resolved.
Beyond the latter distance this typical cluster becomes unresolved
at its center even when observed with {\it HST}.  At the 20~Mpc distance
of The Antennae galaxies, even the {\it half-light} radius of such a
cluster can no longer be resolved with {\it HST} and has to be estimated
by placing an upper limit on it.  Because of these resolution limits, we
first review the fully resolved LPs of young massive clusters within the
Local Group and then the only partially resolved LPs of clusters beyond.

\section{Young Massive Clusters in the Local Group}

\subsection{Magellanic Cloud Clusters}

Much new information on LPs of YMCs in the Local Group has been published
during the past five years, most of it for clusters in the Magellanic
Clouds.  Together the LMC and SMC host an estimated $\sim$6,000 clusters,
most of which are $\la$3~Gyr old, offering us a rich sample of YMCs in the
10$^6$\,--\,10$^9$ yr age range.  Mackey \& Gilmore (2003a,b; hereafter
MG03a,b) have just published the first systematic {\it HST}/WFPC2 study of
LPs of 53 rich clusters in the LMC and 10 in the SMC, yielding profiles
of sub-arcsecond resolution well into the cores ($1\arcsec = 0.24$ pc at
LMC distance).  Some of their main results are as follows.

\begin{figure}[t]
\centerline{\hbox{
\psfig{figure=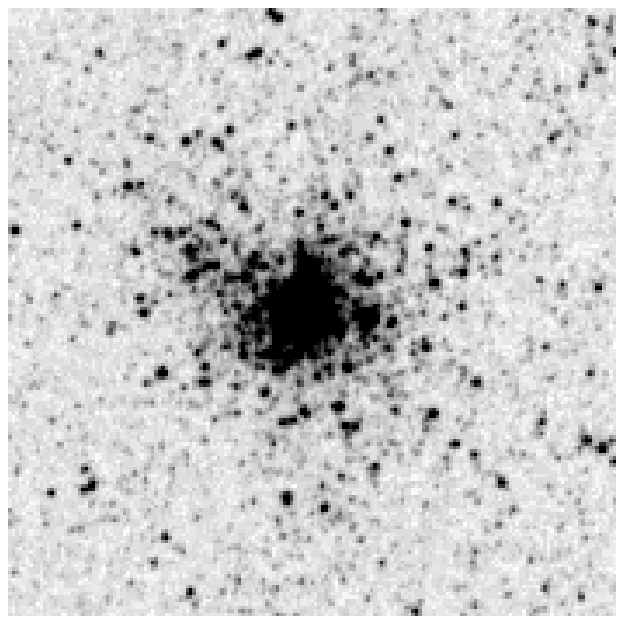,height=5.5cm}
\hskip 0.2truecm
\psfig{figure=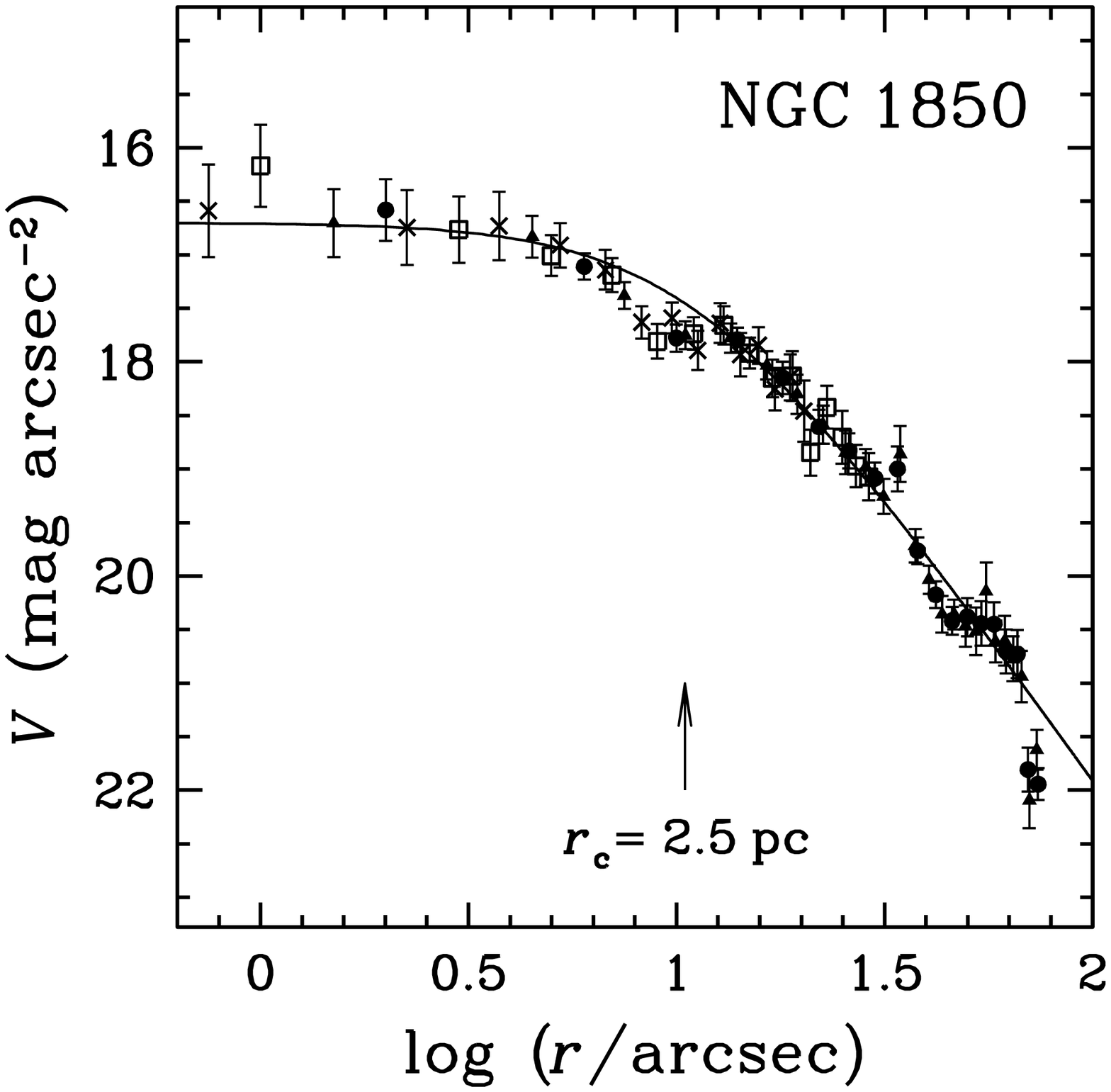,height=5.5cm}
}}
\caption{
The 32\,Myr-old LMC cluster NGC\,1850 ($5\arcmin\times 5\arcmin$ field,
NED/DSS) and its $V$-luminosity profile showing dips and bumps indicative
of clumpy structure (from Mackey \& Gilmore 2003a).
}
\label{fig3}
\end{figure}

In general, virtually all observed profiles of clusters younger than 1~Gyr
are well fit by EFF profiles, thus confirming the findings by EFF87.  The
slopes of the power-law envelopes lie in the range $\gamma = 2.0$\,--\,3.8,
with a median $\gamma \approx 2.6$.  Many profiles show bumps, steps,
and/or wiggles, indicating the presence of significant substructure.  As
an example, Fig.~\ref{fig3} shows the 32~Myr old cluster NGC 1850 and
its profile.  Notice the inner dip at $r \approx 8\arcsec$ and the outer
bump near $r = 60\arcsec$ ($\log r = 1.78$), which is created mainly by the
distinct subclump of stars to the West, sometimes called NGC 1850\,B.
(The slope of the best-fit EFF profile shown in the figure is likely too
shallow because of the limited extent of the WFPC2 data, and
the fitted profile should probably go through the last group of data points.)
Subclumps are seen in the envelopes of most very young clusters and can still
be traced in some clusters several 100~Myr old, suggesting that cluster
formation itself may be hierarchical (Kroupa 1998). Presumably, this
clumpiness accelerates the initial relaxation of newborn clusters.

Interestingly, none of the observed 41 LMC clusters younger than 3~Gyr
shows evidence of having undergone core collapse, as would
be indicated by a profile with power-law shape at the very center (MG03a).
However, such post-core-collapse (PCC) structure is found in $3 \pm 1$ of the
12 observed old ($\ga$10~Gyr) LMC globulars. These and a few other known
PCC clusters have central power-law profiles of slope $\approx -$0.7 and
represent about $20\% \pm 7\%$ of the old cluster population in the LMC,
a similar PCC fraction as is observed among the globulars of the Milky Way.
These observations support the theoretical result that---at least for
single-stellar-mass systems---core collapse takes 12\,--\,19~$t_{\rm rh}$
to occur, or about 300 central relaxation times (e.g., Binney \&
Tremaine 1987).

Correlations between the core radius and various other cluster parameters
have been searched for, yet few have been found.  Specifically, the
core radius does not seem to correlate with cluster mass.  The one
significant correlation found is between the core radius and cluster age,
as first shown by Elson, Freeman, \& Lauer (1989).

\subsubsection{Trends of Core Radius with Age.}

\begin{figure}[t]
\centerline{\hbox{
\psfig{figure=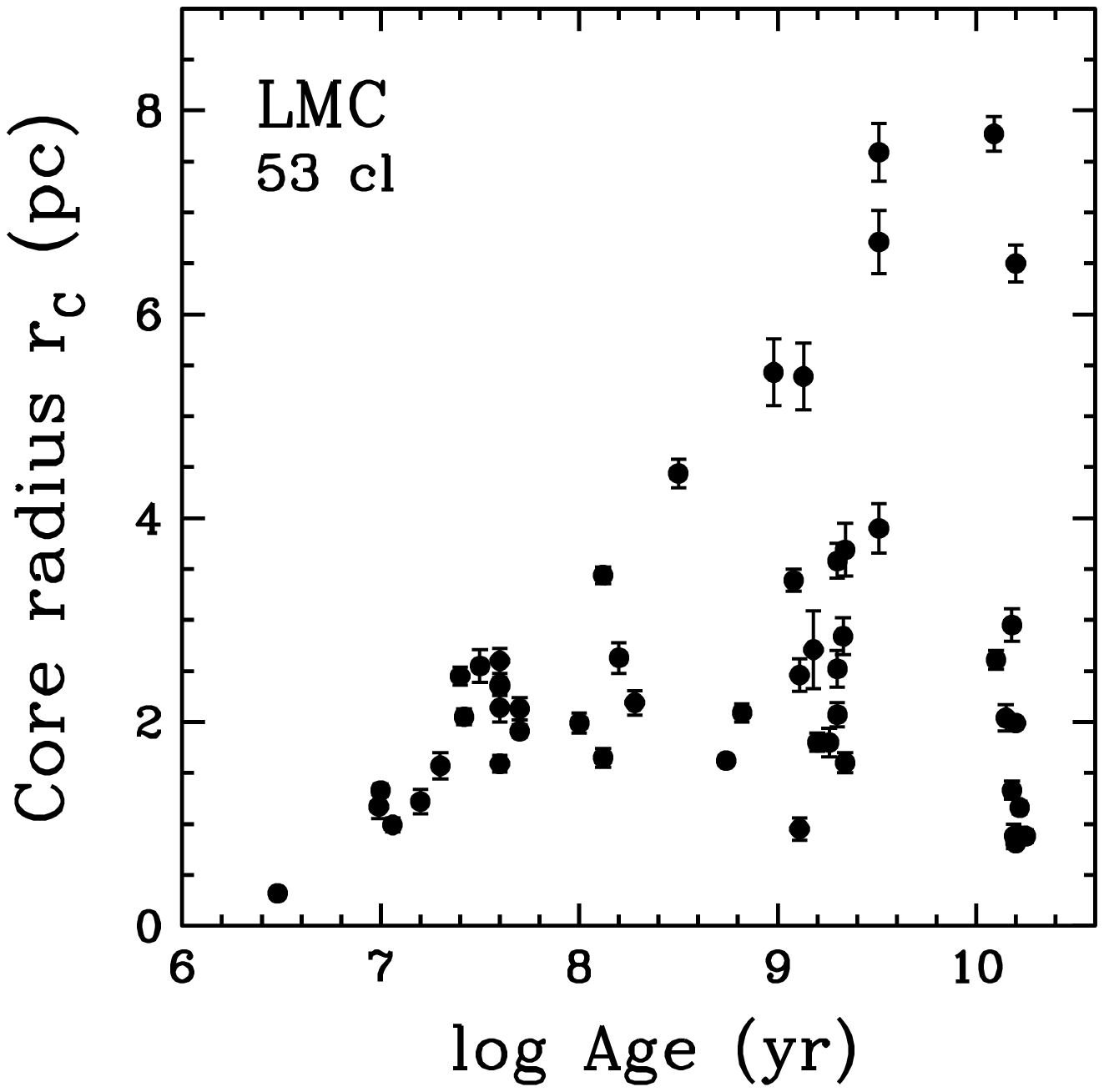,height=5.5cm}
\psfig{figure=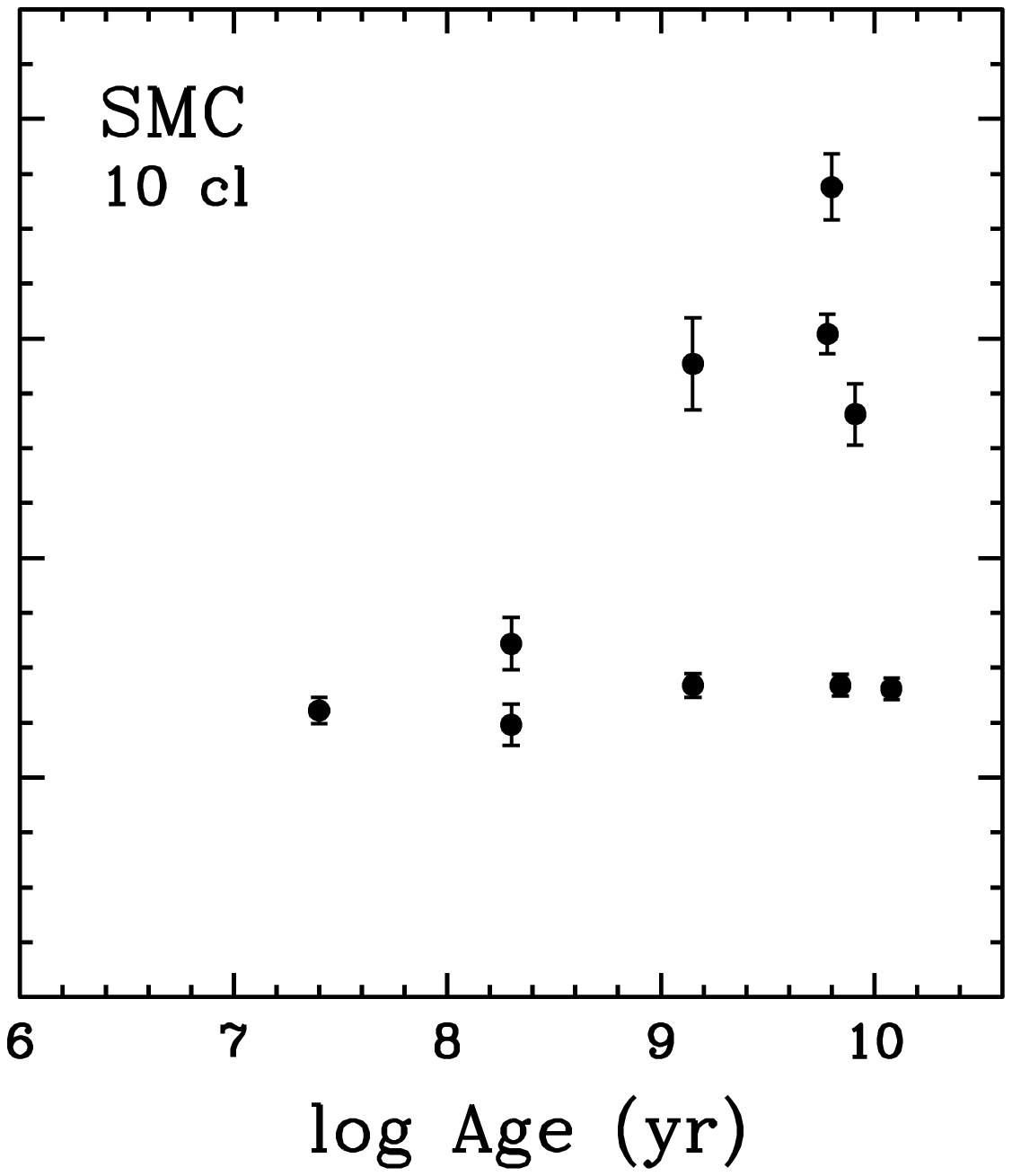,height=5.5cm}
}}
\caption{
Core radii plotted vs.\ age for LMC and SMC clusters (MG03a,b).
}
\label{fig4}
\end{figure}

Figure~\ref{fig4} shows the core radii of the 53 LMC and 10 SMC clusters
observed by Mackey and Gilmore (MG03a,b) plotted versus cluster age $\tau$.
The new {\it HST} data confirm Elson et al's (1989) finding that the core
radii of the youngest clusters are $\la$1~pc and show little scatter, while
those of clusters older than 20~Myr tend to increase with age at least
until $\tau \approx 1$~Gyr and show large scatter at $\tau \ga 10$~Gyr
($0.8 \la r_{\rm c} \la 8$ pc).  However, the relation now appears to 
have {\it two branches}: a lower branch containing about 3/4 of the
total cluster population and reaching a maximum {\it mean} core radius of
$\sim$2.5~pc at $\tau \approx 1$\,--\,2~Gyr before trending toward
smaller $r_{\rm c}$ again, and an upper branch containing about 1/4 of the
clusters and showing core radii that increase in proportion to the
logarithmic age.  As MG03a,b demonstrate, for clusters older than 10~Gyr
as well as for all those older than 1~Gyr the bimodality in the core-radius
distribution is highly significant ($>$99\%, $>$99.5\%), whence the
bifurcation of the $r_{\rm c}$--\,$\log\tau$ relation into two branches
appears real.

The time evolution of core radii traced by the {\it lower branch} is about
as expected: The initial rapid core expansion is likely due to mass
loss from massive stars ($\tau < 1$~Gyr), and is followed by slow
core contraction due to continued energy equipartition and evaporation
of stars, eventually leading to gravothermal instability and core
collapse (Elson et al.\ 1989; see also Binney \& Tremaine 1987).

The time evolution of core radii traced by the {\it upper branch}, however,
remains unexplained.  Clusters of similar age and metallicity on the
upper and lower branches appear to
have had very similar Initial Mass Functions (IMF) down to
$\sim$0.8 $M_{\odot}$, whence IMF variations can be ruled out as an
explanation for the existence of two branches (de Grijs et al.\ 2002b).
$N$-body simulations seem to also rule out as viable explanations any
possible tidal-field variations due to different cluster orbits and
hypothetical large variations in the primordial binary fraction
(Wilkinson et al.\ 2003).  Perhaps most promising is the very recent
suggestion that core formation and expansion may be driven dynamically
by the central accumulation of massive stars and their black-hole
remnants (Merritt et al.\ 2004).  The observed strong segregation of 
massive stars toward the cluster centers (e.g., de Grijs et al.\ 2002a)
and simple model simulations that reproduce the approximately linear
increase of $r_{\rm c}$ with logarithmic age both seem to support this
hypothesis.  Yet, the hypothesis does not explain why the scatter in
core radii increases dramatically with age and why there should be
two branches.

\subsubsection{Newborn Cluster R136/NGC 2070.} 
\begin{figure}[t]
\centerline{\hbox{
\psfig{figure=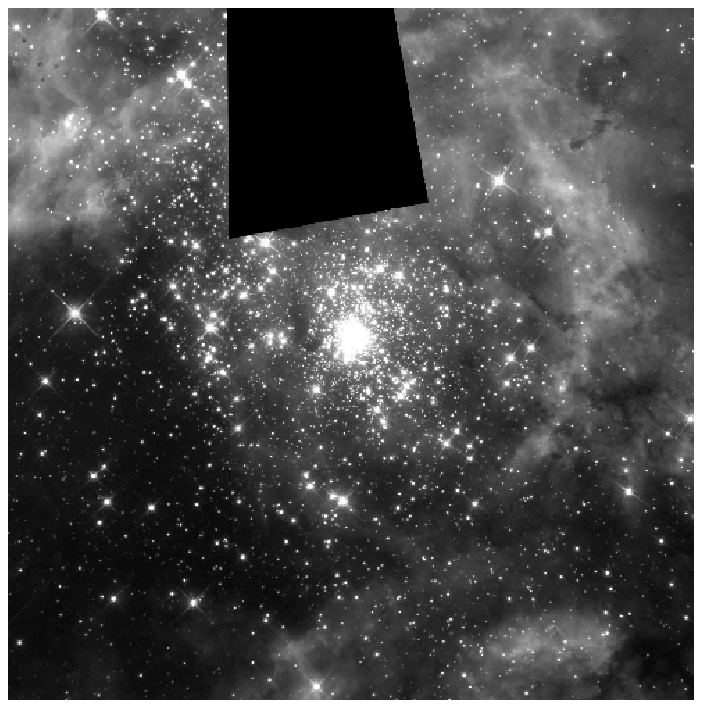,height=5.5cm}
\hskip 0.2truecm
\psfig{figure=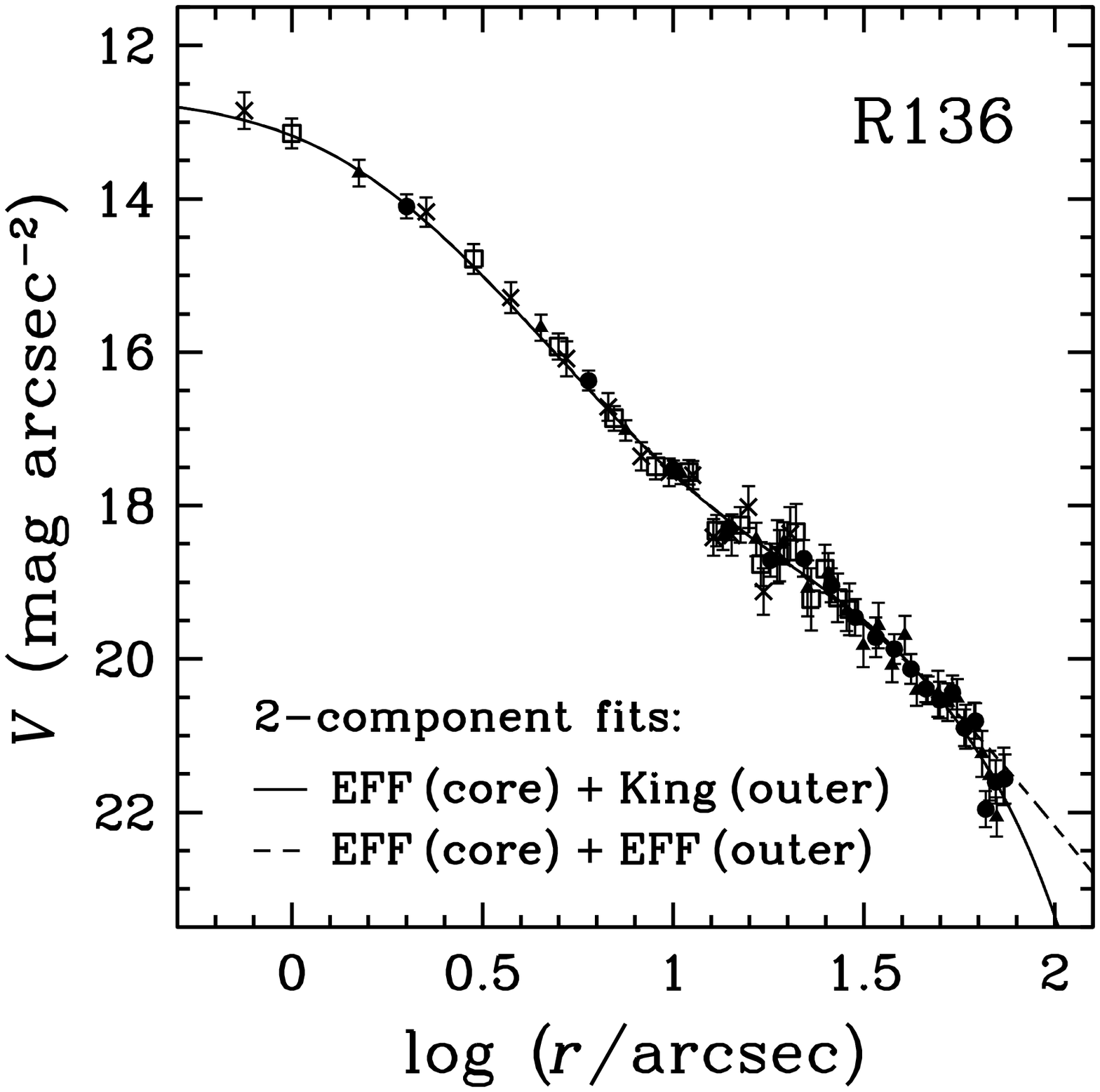,height=5.5cm}
}}
\caption{
Newborn cluster R136/NGC 2070 ($115\arcsec\times 115\arcsec$ field, NASA)
and $V$-luminosity profile with two-component fits superposed (from MG03a).
}
\label{fig5}
\end{figure}

R136 is the
high-surface-brightness core of the 3\,--\,4~Myr old cluster NGC 2070
located at the center of the 30 Doradus nebula, the most luminous H\,II
region in the Local Group.  Figure~\ref{fig5} shows the cluster and its
luminosity profile.  This profile is the only one among the 63 cluster
profiles measured by MG03a,b that clearly shows a two-component structure.
It is well fitted by a central EFF model profile with $r_{\rm c} \la 0.32$ pc
($1\farcs3$) within the core ($r \la 10\arcsec$), and by a second EFF
(or King) profile with $r_{\rm c} \approx 3.7$\,--\,8 pc
(15\arcsec--\,33\arcsec) beyond the core (Meylan 1993; MG03a).

Because R136/NGC 2070 is by far the youngest massive Magellanic Cloud
cluster and possibly a globular cluster in formation (e.g., Kennicutt \&
Chu 1988), its study is of great interest and promises insights into the
earliest dynamical evolution of YMCs.  The evidence for mass segregation
in it is strong: While the core radius is $r_{\rm c} = 0.32$ pc for light
from all stars, it drops to $r_{\rm c} = 0.09$ pc for massive stars of
$>$20 $M_{\odot}$ and $\sim$0.03 pc for those of $>$40 $M_{\odot}$
(Brandl et al.\ 1996).  It is this strong central concentration of the
most massive and luminous stars that led, during the 1980s, to the mistaken
belief that a supermassive star (``R136a'') of $\sim$2500 $M_{\odot}$
might sit at the center of the cluster.

As we now know, there is a strong concentration of O3, O4, and WN stars
at and near the center, some with masses thought to be as high as
130\,--\,150 $M_{\odot}$ (Massey \& Hunter 1998).  Spectroscopy and
IR observations of these and other, pre-main-sequence stars suggest that
lower-mass stars began forming in NGC 2070 about 4\,--\,5~Myr ago, while
the most massive stars formed a mere 1\,--\,2~Myr ago and quenched further
star formation via their strong winds.  Yet, despite this sequential
formation  the {\it overall\,} IMF of NGC 2070 appears to be essentially
normal and
Salpeter-like.  Specifically, searches for low-mass stars show evidence
of rising numbers right into the core down to $\sim$2.8 $M_{\odot}$
(Hunter et al.\ 1995) and even 0.6 $M_{\odot}$ (Sirianni et al.\ 2000).
Although because of the mass segregation the IMF does seem to flatten
below $\sim$2 $M_{\odot}$ within R136 itself, overall it appears very
similar to the Kroupa (2001) or Chabrier (2003) IMFs now generally
thought to be characteristic of star formation in the Milky-Way disk and
in loose young clusters. Hence, the observations of this newborn cluster
clearly contradict some theoretical predictions that the IMF of stars
formed in high-density regions should be very top heavy.  They also raise
doubts about the importance of occasionally-claimed primordial mass
segregation.

An interesting question is whether the unusual core-within-a-core structure
of R136/NGC 2070 may have led to early core collapse.  Spitzer (1969)
showed that under certain circumstances clusters with a wide range of
stellar masses may experience an {\it equipartition instability}, in which
the most massive stars form a dense subsystem at the core of the lighter
stars.  While exchanging energy with the lighter stars, this subsystem
can evolve away, rather than toward, equipartition, contracting rapidly
and leading to accelerated core collapse. Whether this did or did not
happen in R136 depends critically on the core radius measured as a function
of mass and on the estimated central relaxation time.  MG03a find marginal
evidence for a central power-law cusp of slope $-$1.17, but conclude---like
Brandl et al.\ (1996) before---that the evidence favors strong and rapid
dynamical mass segregation over any post-core-collapse state.

In short, the two-component profile of this $\sim$3\,--\,4~Myr old cluster
suggests that (i) strong dynamical evolution and mass segregation occur
very early on and (ii) core growth during the first $\sim$10~Myr may be
complex.

\subsection{Clusters in Other Local Group Galaxies}

Surprisingly little information exists on the luminosity profiles of
YMCs in Local Group galaxies other than the Magellanic Clouds.  For
the Milky Way, M31, and M33 the situation is as follows.

\subsubsection{Milky-Way Clusters.} The Milky Way seems to harbor
few YMCs comparable to those in the LMC and SMC.  Attempts to measure
their LPs suffer from severe disk-star contamination and extinction.
Perhaps the best studied is the 1\,--\,2~Myr old cluster NGC 3603,
whose $A_V = 4.5$~mag is relatively benign. It is one of the most massive
young clusters known in the Milky Way, yet it has only about 1/40th the
mass of R136/NGC 2070.  Like the latter, it features low-mass stars
of 0.1\,--\,1 $M_{\odot}$ right into its center and has a normal
IMF (e.g., Brandl et al.\ 1999).  Infrared observations yield a LP
with a core radius of $r_{\rm c} = 0.78$ pc (23\arcsec) and extending
out to at least 5 pc, where it becomes lost in the light of foreground
stars (N\"urnberger \& Petr-Gotzens 2002).  The situation is even more
challenging for the Milky Way's central YMCs, with the Arches cluster
showing extinction {\it variations} of $\Delta A_V \approx 10$ mag over
15\arcsec\ (Stolte et al.\ 2002) and, thus, offering little hope for any
luminosity profiles in the near future.

\subsubsection{M31 Clusters.}  Although some YMCs resembling young
globulars have long been known to exist in M31 (e.g., van den Bergh 1969),
none have had their LPs measured.  Yet, at least for the four
60\,--\,160~Myr old YMCs observed by Williams \& Hodge (2001) with
{\it HST}/WFPC2, LPs would be easy to derive from the archival images.
Luminosity profiles have, however, been measured for {\it old} massive
clusters and show evidence for PCC structure and extended envelopes
(Grillmair et al.\ 1996) and, in the still controversial case of Cluster G1,
for a central black hole of mass $2.0^{+1.4}_{-0.8}\times 10^4 M_{\odot}$
(Gebhardt, Rich, \& Ho 2002).

\subsubsection{M33 Clusters.}  M33\ hosts a rich system of
$10^6$\,--\,$10^{10}$~yr old clusters similar to those in the Magellanic
Clouds.  For $\sim$60 of these clusters, LPs have been derived from
{\it HST}/WFPC2 images and fitted with King profiles (Chandar, Bianchi,
\& Ford 1999).  The measured core radii are $r_{\rm c}\approx
0.2$\,--\,2 pc, and there is evidence for extended envelopes.  At least
for clusters younger than $\sim$3~Gyr, the analysis needs to be repeated
with the more appropriate EFF profiles.

\section{Young Massive Clusters Beyond the Local Group}

Even with {\it HST}, the cores of YMCs become marginally resolved
around 2\,--\,4~Mpc and unresolved beyond (Fig.~\ref{fig2}).  Special
software has been developed to analyze observations of partially resolved
clusters by fitting King or EFF model light distributions (Larsen
1999; Carlson \& Holtzman 2001).  Therefore, we can still extract some
size and shape parameters from YMCs out to at least 20 Mpc.

\begin{figure}[t]
\centerline{\hbox{
\psfig{figure=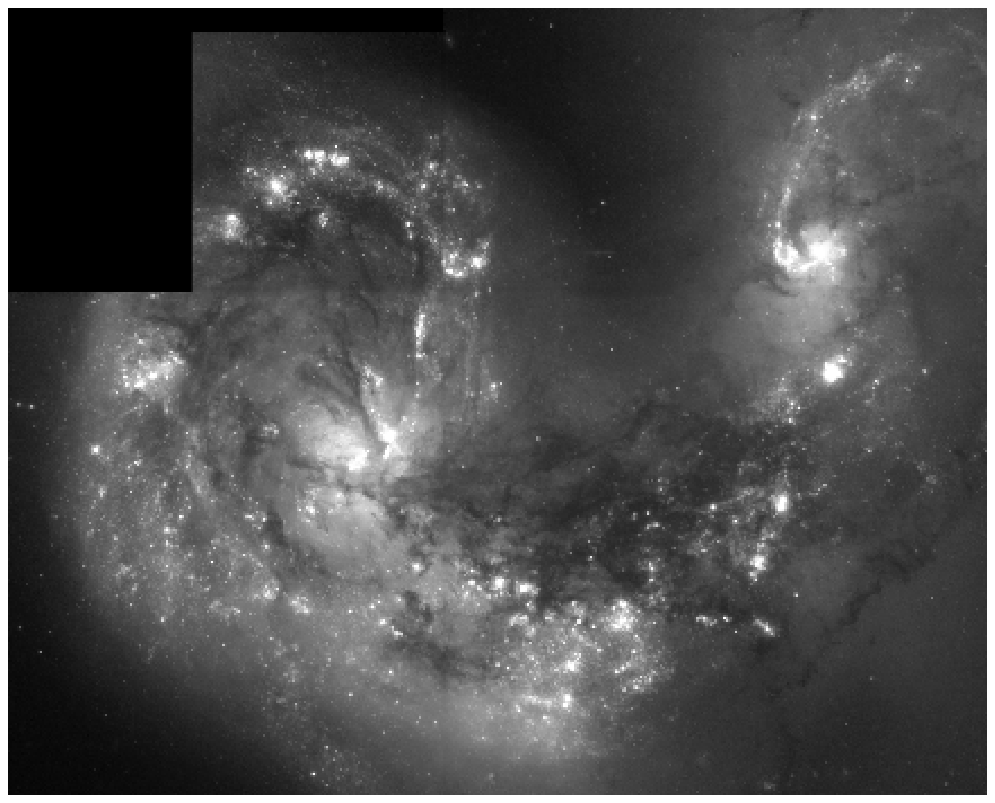,height=5.8cm,angle=270}
\hskip 0.5truecm
\psfig{figure=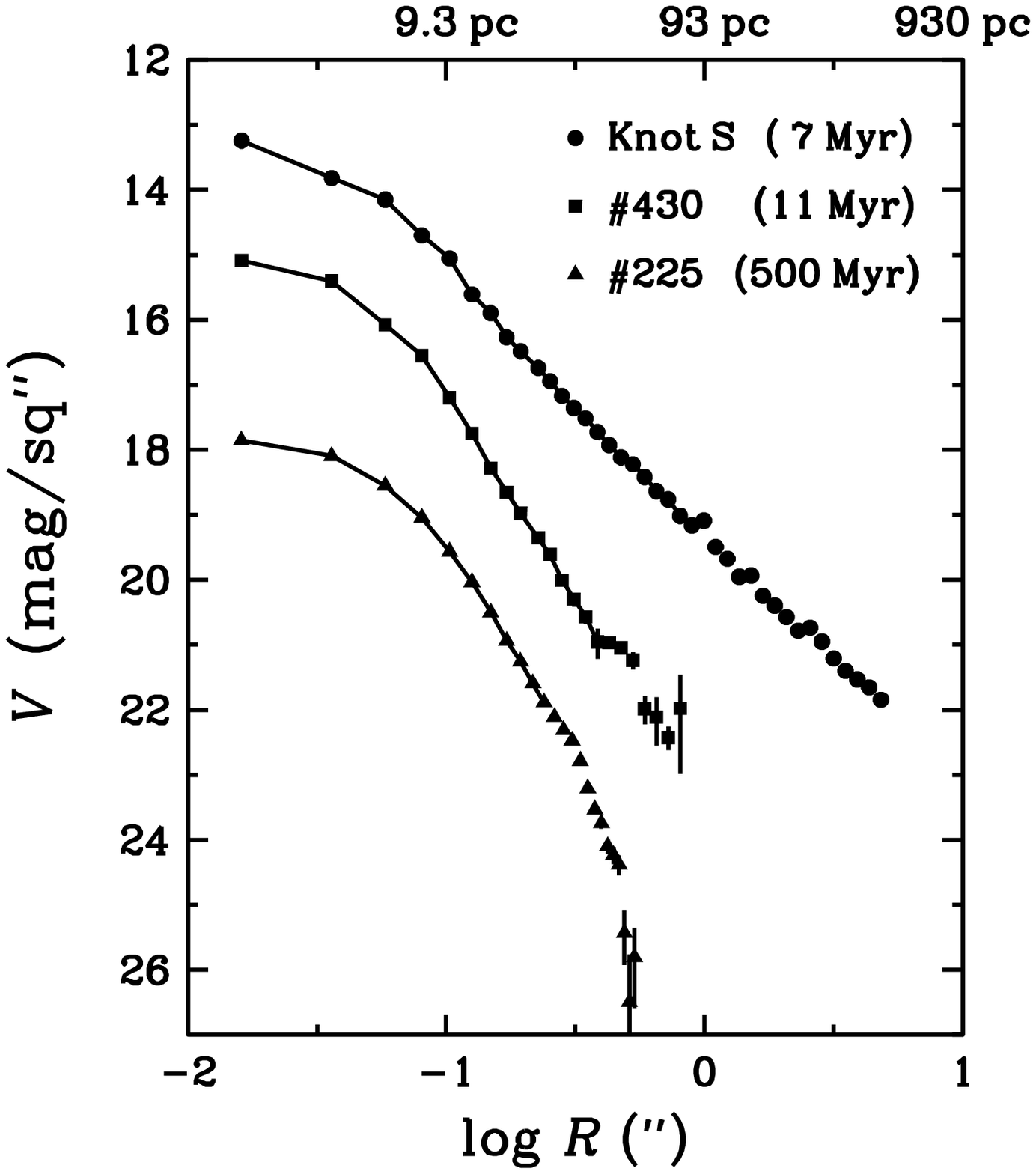,height=5.8cm,angle=0}
}}
\caption{
The Antennae galaxies (NGC 4038/39), and luminosity profiles of three
young massive clusters in them (from Whitmore et al.\ 1999).
}
\label{fig6}
\end{figure}

Among relatively nearby (2\,--\,6 Mpc) galaxies, luminosity profiles have
been measured for YMCs in NGC 1569 (Hunter et al.\ 2000), M82 (de Grijs et
al.\ 2001), and NGC 6946 (Larsen et al.\ 2001). In general, the core radii
of these YMCs are similar to those measured in LMC clusters.
Also, all YMCs younger than $\sim$10$^8$\,yr show power-law envelopes, and
many show clumpy substructure.

An especially interesting case are the 43 globular clusters profiled in
NGC 5128 (4 Mpc, Harris et al.\ 2002), where six of the most luminous
clusters ($-10\la M_V \la -11.3$) show envelopes extending beyond the
best-fit King models.  Although Harris et al.\ interpret these envelopes
as being either due to evaporative mass loss or the remains of stripped
former dwarf galaxies, a third possibility needs to be considered:
The presence of intermediate-age globular clusters in NGC 5128 (Peng, Ford,
\& Freeman 2004) and the high luminosity of the six globulars may indicate
that these are relatively young clusters with incompletely stripped remains
of their initial power-law envelopes.

Even for galaxies more distant than 10 Mpc, there is still much to
learn from luminosity profiles of YMCs.  Especially intriguing are the
hints of dynamical evolution seen among clusters of different ages in
The Antennae (Whitmore et al.\ 1999).  Figure~\ref{fig6} shows the LPs
of three massive clusters there: Knot~S and \#430 are both very
young (7 and 11~Myr), have unresolved cores, and display pure power-law
envelopes like young LMC clusters.  In contrast, the 500~Myr-old
Cluster \#225 shows both a larger, partially resolved core ($r_{\rm c}
= 5.6$ pc) and an envelope with a distinct tidal cutoff.  This suggests
that, like elsewhere, young clusters in The Antennae form with power-law
envelopes, but then get truncated over the next several 100~Myr by
external tidal forces.

Finally, even at the 66 Mpc distance of the merger remnant NGC 7252
evidence for extended envelopes of YMCs can be found.  Among the
$\sim$300 young halo globulars of solar metallicity and ages
300\,--\,600 Myr (Miller et al.\ 1997; Schweizer \& Seitzer 1998), the
five most luminous clusters feature extended envelopes reaching well past
100~pc radius.  The most luminous, NGC\,7252:\,W3 ($M_V= -16.2$), is a
true supercluster with $r_{\rm eff}= 17\pm 2$~pc and a
record-beating mass of $M= (8\pm 2)\times 10^7 M_{\odot}$, corresponding
to 15\,--\,20$\times$ the mass of $\omega$\,Cen (Maraston et al.\ 2004).
Apparently, while orbiting in the halo of NGC 7252 this heavy-weight young
globular of age 300\,--\,500~Myr may have managed to hang on---so far---to
most of its original power-law envelope.

\acknowledgements
I thank Jon Holtzman, Robert Lupton, Dougal Mackey, and Brad Whitmore for
their kind permission and help in reproducing figures, and acknowledge
partial support from the NSF through Grant AST--02\,05994.


\end{document}